\documentclass[twoside]{ilcws10}
\usepackage[latin1]{inputenc}
\usepackage[dvips]{graphicx,epsfig,color}
\usepackage{wrapfig,rotating}
\usepackage{amssymb,amsmath,array}

\begin{document}
\title{
Progress towards a Technological Prototype for a Semi-Digital Hadron Calorimeter based on Glass RPCs} 
\author{N. Lumb$^*$ for the CALICE collaboration
\vspace{.3cm}\\
$^*$Institut de Physique Nucléaire de Lyon \\
69622 Villeurbanne - France
}

\maketitle

\begin{abstract}
The semi-digital Hadronic calorimeter using GRPC as sensitive medium is one of the two options the ILD collaboration is considering for the ILD detector final design.
A prototype of 1~m$^3$ has been conceived within the CALICE collaboration in order to validate this option. The prototype is intended to be as close as possible to the one proposed in the ILD LOI. A first unit of  1~m$^2$ GRPC of 3 mm thickness and fully equipped with a semi-digital electronics readout and new gas distribution design was produced and successfully tested. In 2010 we intend to produce 40 similar units to be inserted in a self-supporting mechanical structure. The prototype will then be exposed to test beams at CERN or at Fermilab for final validation.
\end{abstract}

\section{Introduction}
The Videau concept for a semi-digital HCAL for the ILD detector at the International Linear Collider has been described in previous papers~\cite{tipp09}.  This HCAL consists of stainless steel absorbers with glass RPCs as active layers.  In order to demonstrate the validity of the concept, a technological prototype based on one module of one wheel of the detector is being designed and built.  The technological prototype will be a simplification of a real module in that the maximum RPC size will be 1~m x 1~m (instead of 3~m x 1~m) and only 40 detection layers will be built (instead of 48).  The main challenges are:

\begin{itemize}
\item Detector + electronics thickness $<$ 6mm
\item Minimize dead zones
\item Homogeneous RPC gain
\item Chamber efficiency $>$90$\%$ + minimize multiplicity
\item Full electronics with power pulsing
\item Realistic support structure for absorbers + RPCs
\end{itemize}

In this paper we describe progress towards the goal of constructing the technological prototype, both from the point of view of chamber optimization and in terms of the wider systems aspects.

\section{GRPC developments}

\subsection{Chamber design}
\label{chamber_design}
\begin{figure}[h]
\centerline{\includegraphics[width=1.00\columnwidth]{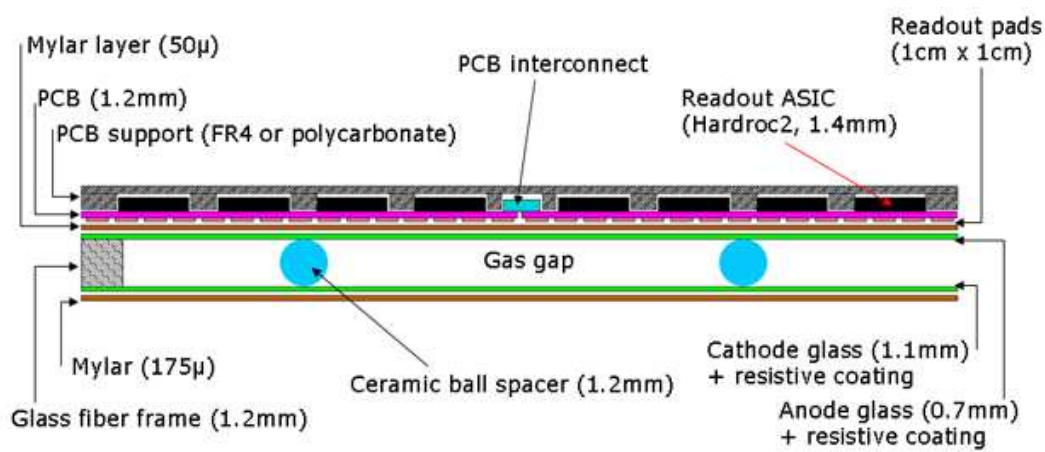}}
\caption{Cross-section through a 1~m$^2$ chamber.}\label{chamber}
\end{figure}

The 1~m$^2$ chambers consist of the elements shown in Fig.~\ref{chamber}.  Precision ceramic balls of diameter 1.2~mm are used as spacers to separate the glass electrodes of thickness 0.7~mm (anode) and 1.1~mm (cathode).  The gas volume is closed by a glass fiber frame.  Read-out pads of area 1~cm x 1~cm are isolated from the anode glass by a thin Mylar foil.  These pads are etched on one side of a PCB; on the other side are located the front-end read-out chips (Hardroc 2b~\cite{hardroc}).  Finally, a polycarbonate spacer (`PCB support' in Fig.~\ref{chamber}) is used to `fill the gaps' between the read-out chips and to improve the overall rigidity of the detector / electronics `sandwich'.  The total theoretical thickness of the assembly is 5.825~mm.  Taking into account air gaps and engineering tolerances, the true thickness is expected to be of the order 6.15~mm.

\subsection{Resistive coating}
\begin{figure}[h]
\centerline{\includegraphics[width=1.00\columnwidth]{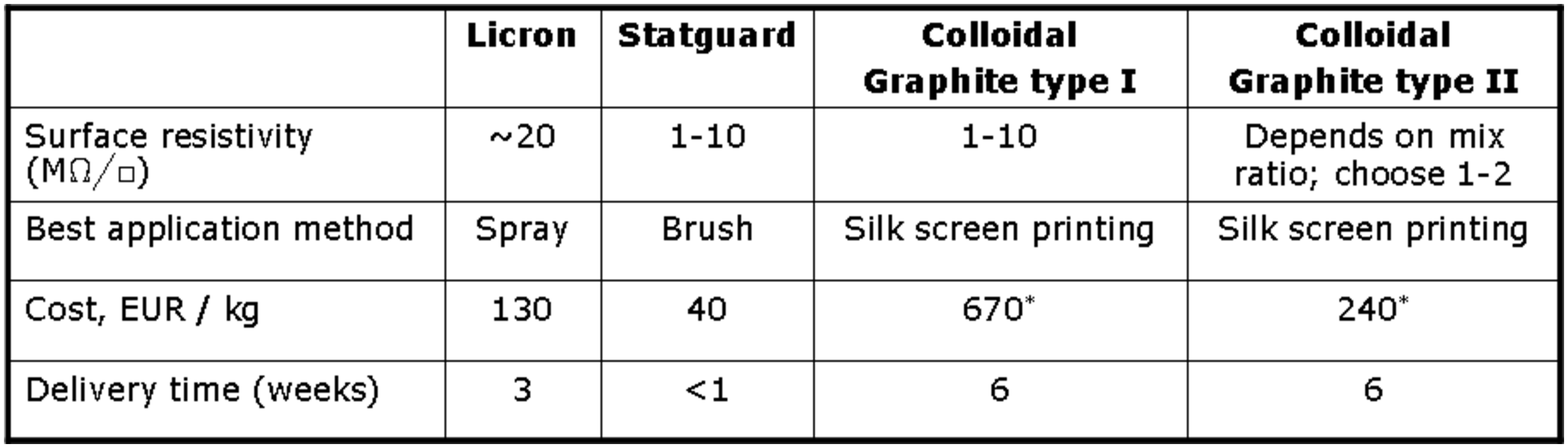}}
\caption{Comparison of commercial products considered for resistive coating.  ($^*$Estimate 20~m$^2$ (10 chambers) / kg using silk screen printing technique).
}\label{coating_table}
\end{figure}

Considerable research effort has been invested in identifying the best resistive coating for the chamber glass.  Fig.~\ref{coating_table} compares some of the products recently studied.

\begin{wrapfigure}{r}{0.5\columnwidth}
\centerline{\includegraphics[width=0.65\columnwidth]{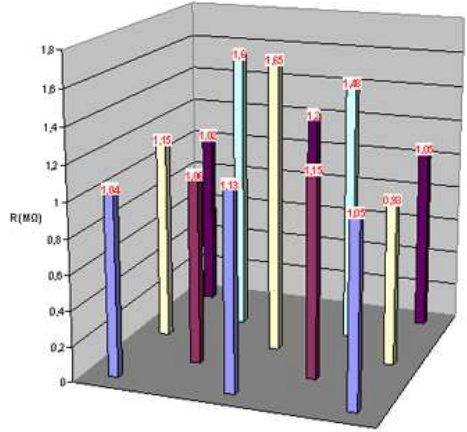}}
\caption{Surface resistivity for 1~m$^2$ glass coated with bi-component colloidal graphite.}\label{colloid_resistivity}
\end{wrapfigure}

The `Licron' product was found to be problematic in that it has the tendency to `migrate' away from the high voltage strip glued to the cathode glass, resulting in loss of the HV contact after a very short time (a few days of continuous use).  We have constructed several large-area GRPCs using the `Statguard' product, an inexpensive floor paint used in ESD applications.  These chambers have been successfully operated (see \cite{tipp09}); however the paint was applied to the glass using a paint brush, which is considered not suitable for mass production.  Attempts to coat large areas with Statguard using the silk screen technique produced unsatisfactory results.  This product also has a long time constant to reach a stable surface resistivity (typically two weeks).  Research was therefore carried out to find a product specifically designed for silk screen printing with the correct surface resistivity.  Two products were identified, both of which are based on colloids containing graphite.  One of these products is a single component paint with a dry surface resistivity of 1-10 M$\Omega/\square$ after being deposited using the silk screen method.  The second product comes as two components which must be mixed by the user.  The surface resistivity may be adjusted over a wide range by changing the mix ratio.  Both products require baking at around 170$^o$C to attain a stable surface resistivity.

The measured surface resistivities at various points over a 1~m$^2$ glass coated with the bi-component paint are shown in Fig.~\ref{colloid_resistivity}.  The mean value is 1.2~M$\Omega$/$\square$ and the ratio of the maximum to minimum values is 1.8.  A study was also made of the repeatability of the surface resistivity between different mix batches.  It was found that surface resistivities in the range 1-2~M$\Omega$/$\square$ could be reliably reproduced, again with a factor of approximately two between minimum and maximum values.  These results are considered entirely satisfactory, since it is known from previous tests~\cite{tipp09} that significant impact on chamber performance begins to be measurable only for resistivity variations of the order of a factor 10.

\subsection{Distribution of gas within the chamber}

\begin{figure}
\centerline{\includegraphics[width=1.00\columnwidth]{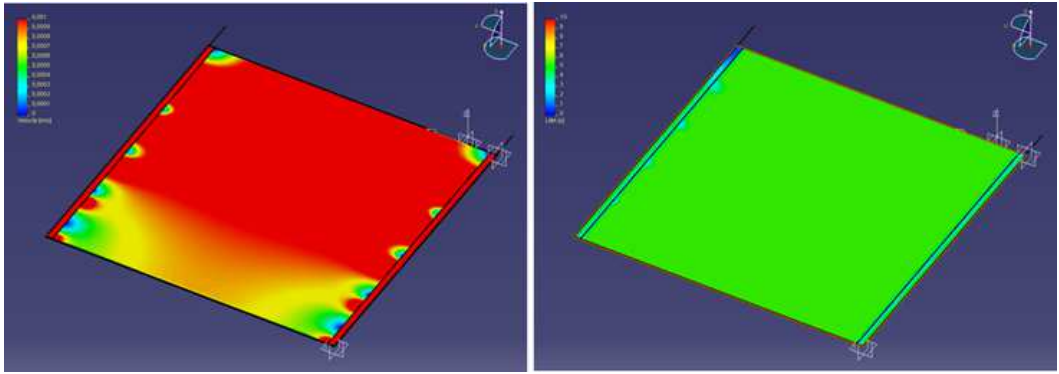}}
\caption{(a) Left:  gas speed profile in the range 0-1~mm/s;  (b) Right:  Least mean age profile in the range 0-10~s.}\label{gas_distribution}
\end{figure}

Gas distribution within the RPC is improved by channeling the gas along one side of the chamber and releasing it into the main gas volume at regular intervals.  A similar system is used to collect the gas at the other side of the chamber.  A finite element model has been established to verify the gas distribution - see \cite{tipp09} for details.  Recent results from this model are shown in Fig.~\ref{gas_distribution}.  The simulation confirms that the gas speed is reasonably uniform over most of the chamber area and that this design significantly improves the distribution of gas with respect to a chamber with no gas channels.  A profile of the quantity known as the `Least Mean Age' (LMA) was also generated by the model.  This is defined as the time for gas to reach a given point in the chamber after entering the volume, including the effects of diffusion.  At most points in the chamber the LMA was calculated to be around 5~s, indicating that diffusion probably plays an important role in the distribution of the gas.

\subsection{Read-out electronics}

\begin{wrapfigure}{r}{0.57\columnwidth}
\centerline{\includegraphics[width=0.45\columnwidth]{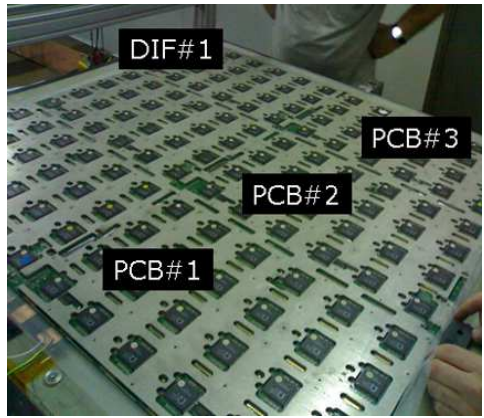}}
\caption{Fully assembled 1~m$^2$ chamber showing read-out electronics.}\label{electronics}
\end{wrapfigure}

The PCBs supporting the read-out chips are currently limited in size so the 1m$^2$ area is covered using six PCBs of area 500 x 333~mm.  The PCBs are linked electrically by micro-soldering or (next generation) using special low-profile connectors.  Each PCB supports 24 ASICs, with each ASIC having 64 read-out channels, for a total of 144 ASICs and 9216 channels per detector.  The total number of channels for the technological prototype will therefore be 368,340.  Fig.~\ref{electronics} shows a fully assembled square meter of electronics.  Note the special `mask' filling the gaps between ASICs, as mentioned in section \ref{chamber_design}.
\vspace{5cm}

\pagebreak

\section{Technological prototype: systems aspects}

\subsection{The RPC cassette and mechanical superstructure}

\begin{wrapfigure}{r}{0.5\columnwidth}
\centerline{\includegraphics[width=0.45\columnwidth]{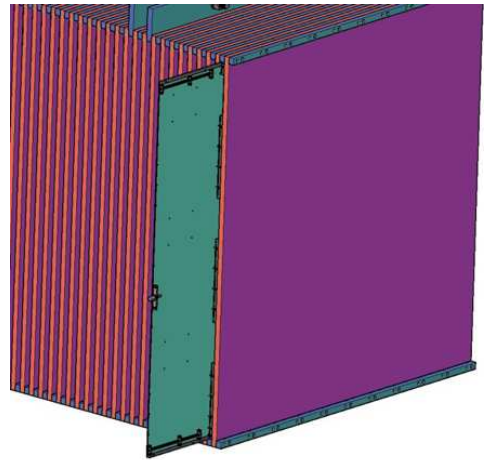}}
\caption{The mechanical super-structure showing one cassette inserted.}\label{superstructure}
\end{wrapfigure}

The GRPC and its associated electronics are housed in special cassettes which protect the chambers and ensure that the read-out boards are in intimate contact with the anode glass.  The cassette is a thin box consisting of 2~mm and 3~mm thick stainless steel plates separated by stainless steel spacers which form the walls of the box.  These spacers are precision machined such that the upper metal plate just touches the PCB support (see Fig.~\ref{chamber}).  The cassette forms part of the absorber layer and the thickness of the absorber plates is 15~mm, giving a total of 20~mm per layer.

The RPCs and their cassettes will be produced independently of the mechanical super-structure containing the absorber plates.  The plates will be assembled horizontally with spacers to provide the gaps into which the cassettes are inserted - see Fig.~\ref{superstructure}.  Recent tests have validated the insertion technique and allowed us to specify tolerances for the final assembly.

\subsection{Thermal modeling}

\begin{wrapfigure}{r}{0.5\columnwidth}
\centerline{\includegraphics[width=0.45\columnwidth]{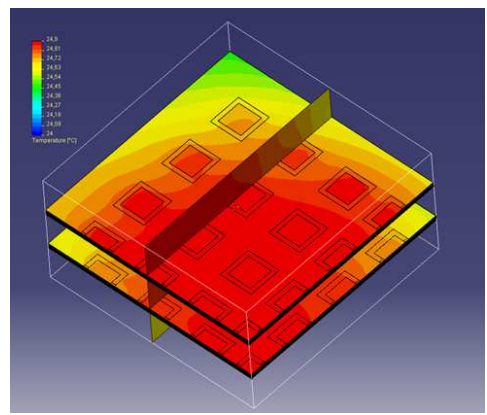}}
\caption{Finite element model of thermal behaviour of one quarter of a two detector layer system.}\label{thermal_results}
\end{wrapfigure}

The technological prototype will feature a high number of electronics channels, the heat from which must be evacuated.  Power pulsing of the electronics will in principle ensure that the power per channel is extremely low ($\approx$10~$\mu$W); nevertheless a thermal model has been developed to study the worst case scenario of no power pulsing.  In this scenario, the power consumption is a factor 100 higher.  Two detector planes with electronics and three absorber layers were modeled.  Based on symmetry arguments, it was considered necessary to simulate only one quarter of the full area of each layer - see Fig.~\ref{thermal_results}.

Cooling was assumed to be entirely passive, i.e. by convection only, for an ambient temperature of 20$^o$C.  Under these conditions, the model predicts a maximum rise in temperature of 5$^o$C.  Active cooling is therefore considered unnecessary in the present design.

\subsection{High Voltage system}

\begin{wrapfigure}{r}{0.5\columnwidth}
\centerline{\includegraphics[width=0.45\columnwidth]{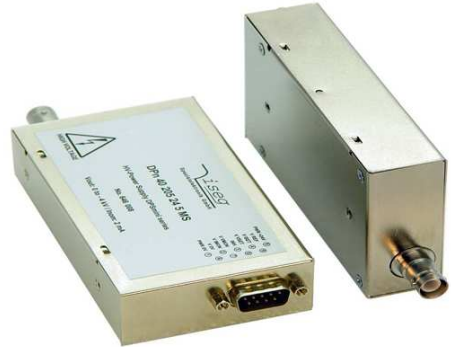}}
\caption{Candidate HV module from Iseg based on the Cockcroft-Walton voltage multiplier.}\label{iseg_hv}
\end{wrapfigure}

The high voltage system will be based on the Cockcroft-Walton voltage multiplier.  Such a system allows low voltage (0-5~V) cables to be used up to the cassettes, greatly reducing the volume of cable insulation inside the detector.  A suitable module is being developed in collaboration with Iseg - see Fig.~\ref{iseg_hv}.

The module, one for each RPC, will deliver up to 10~$\mu$A with a voltage range of 0-10~kV.  In addition, it must be sufficiently low-profile ($<$~24~mm) to fit between cassettes.  Control and monitoring of voltages will be available via ethernet link.

\subsection{Gas distribution}
\begin{figure}[h]
\centerline{\includegraphics[width=1.00\columnwidth]{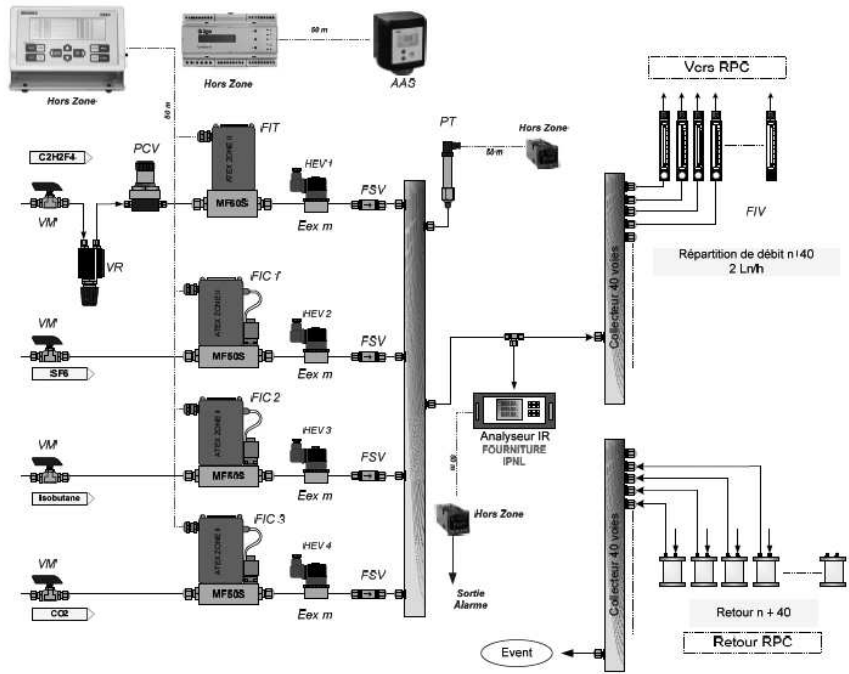}}
\caption{Forty channel gas distribution system for the technological prototype.}\label{gas_system}
\end{figure}

A gas mixture based on tetrafluoroethane (C$_2$H$_2$F$_4$) will be used, with SF6 and isobutane quenchers.  A 40-channel gas distribution system has been ordered - see Fig.~\ref{gas_system}.

Precision mass flow meters are used to regulate the flow of each gas component to within $\pm$2$\%$.  The gases are mixed and then split off into 40 parallel outlets.  The flow from each outlet can be independently regulated and monitored using rotameters.  Each of the 40 return lines is equipped with a bubbler.  The system is certified for use in flammible environments to ATEX zone 2 level.

\section{Conclusions}

The construction of 1~m$^2$ GRPCs with good detector performance is well understood.  Prototype chambers with uniform resistive coatings, constant gas gap and optimized gas distribution have been produced.  The development of suitable read-out electronics based on the Hardroc 2 is chip well advanced.
Special protective cassettes for the chambers have been designed, constructed and validated.
The mechanical super-structure of the technological prototype is at an advanced design stage and tenders for purchase of materials for this important element will be sent out soon.
An insertion test has been performed to validate the technique for placing the cassettes between the absorber plates.
A thermal simulation of the system has been completed and the the passive cooling concept has been confirmed.
Design and testing in collaboration with industry of a multi-channel voltage multiplier system is well advanced and a forty channel gas system has been ordered.

\section{Bibliography}


\begin{footnotesize}


\end{footnotesize}


\end{document}